# Study on FLOWSIM and its Application for Isolated Signalized Intersection Assessment

*Yuhan JIA*[1,2], *Jianping WU*[1,2], *Yiman DU*[1,2], *Geqi QI*[1,2]

(1.Department of civil engineering, Tsinghua University, Beijing, 100084, China; 2.Jiangsu Province Collaborative Innovation Center of Modern Urban Traffic Technologies, Nanjing, 210096, China)

**Abstract**：Recently the traffic related problems have become strategically important, due to the continuously increasing vehicle number. As a result, microscopic simulation software has become an efficient method in traffic engineering for its cost-effectiveness and safety characteristics. In this paper, a new fuzzy logic based simulation software (FLOWSIM) is introduced, which can reflect the mixed traffic flow phenomenon in China better. The fuzzy logic based car-following model and lane-changing model are explained in detail. Furthermore, its applications for mixed traffic flow management in mid-size cities and for signalized intersection management assessment in large cities are illustrated by examples in China. Finally, further study objectives are discussed.

**Key words**： microscopic simulation; FLOWSIM; mixed traffic flow; intersection management assessment

## 1 Introduction

Recently traffic problems are becoming increasingly serious in urban area such as traffic jams, road accidents and vehicle emission. Among them, traffic congestions affect urban citizens' daily life severely, especially at intersections. Actually congestion not only enlarges the travel cost, but also produces more emission to environment. Typically the construction of traffic facilities like building expressways and enlarging intersections can solve this problem. However, given the current condition, in most cities there is no room for road widening. Thus a common sense has been reached that advanced traffic control methods are the key point to deal with traffic problems above. An important part of traffic control is the intersection management, including signal control and channelization optimization. As a result, many traffic investigators have been focusing on the methods to improve intersection management.

From last century, traffic engineers have developed a series of microscopic traffic simulation software. The theoretical bases are car-following theory and lane-changing theory [1]. Several microscopic softwares are widely used like AIMSUN form Spain, TRANSIMS form USA, VISSIM form Germany and FLOWSIM form UK [2,3]. Compared with traditional management methods, simulation software can provide optimization plans by computing without large-scale field measurement. This can solve the issue that it is impossible in urban area in large cities to adjust real control plans for evaluation. With simulation software, traffic researchers can study signal optimization [4], road reconstruction [5], intersection management [6], and public traffic system [7].

This paper focuses on the software of FLOWSIM and its application in China, for both mid-sized cities and large cities. The software is briefly introduced in section 2 with the demonstration of software interface. In section 3 the theoretical fundamental of fuzzy logic is given for deeper understanding, which consists of car-following theory and lane-changing theory. Then the applications of FLOWSIM for mixed traffic flow management and real time signalized intersection assessment are described. Finally, the last section is the summary and further study plan.

## 2 Software of FLOWSIM

FLOWSIM is developed with Fuzzy Logic based Motorway traffic Simulation Model from University of Southampton. Typical simulation models use a deterministic way to describe driver's behavior which may cause inaccuracy form reality. In FLOWSIM, fuzzy logic is used as a kernel theory to build car-following model and lane-changing model because it can reflect the drivers' stochastic behavior objectively. FLOWSIM was mainly calibrated in UK as well as some other European countries and showed good accuracy for practical application in many projects. From 2000 the research and development team began to collect data and adjust model in China and now this simulation software has been put into use in many domestic cities.

FLOWSIM can be applied for urban road system, artery,

highway, and the combination of them. Common sceneries can be simulated like signalized or unsignalized intersection control, ramp metering and roundabout management. It can also be used for the influence analysis like traffic accidents, traffic signs and extreme weather. The highlight is that FLOWSIM can realize the simulation of pedestrian and bicycle flow corresponding to the real traffic state in China [2]. Fig. 1 shows the simulation interface of an intersection. It is observed that there are many conflicts between vehicles and non-motor vehicles.

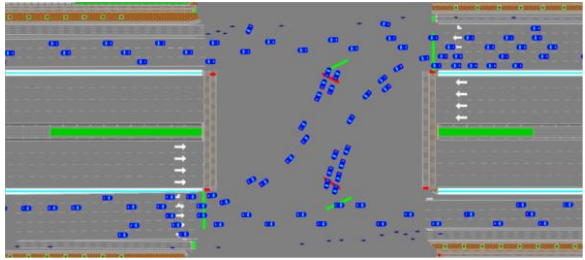

Fig. 1 Simulation surface of FLOWSIM.

FLOWSIM offers the graphical toolbox for the visualization of traffic engineering. Road and intersection can be truly reproduced in modeling surface and several formats of picture can be imported as the background layer. Furthermore there are related tools to edit the properties of road like its grade, turning curvature, maximum speed and capacity. For intersection modeling, traffic signals and road channelization are also provided for the operational performance evaluation of different plans.

After road modeling and data input, FLOWSIM can run the simulation for a user-defined time. Typically we can enter the existing traffic state into software and find out the shortcomings by examining the simulation results like travel time and queuing length. This process provides a method to assess the intersection control quality and make further optimization without too much workload on the spot.

Simulation software is often used for the comparison of optimized and existing traffic management plan. In FLOWSIM several assessment criteria are offered like delay time, travel time, road capacity and queuing length. In addition, FLOWSIM provides a three-dimensional animation for an intuitive understanding, which can be shown in Fig. 2. Also, other applications such as protected area feedback control, multiply OD matrix input and intersection feedback management can also be realized.

In summary, the main functions and operational process can be shown in Fig. 3.

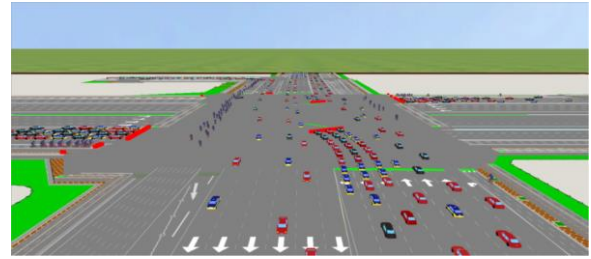

Fig. 2 Three-dimensional simulation in FLOWSIM.

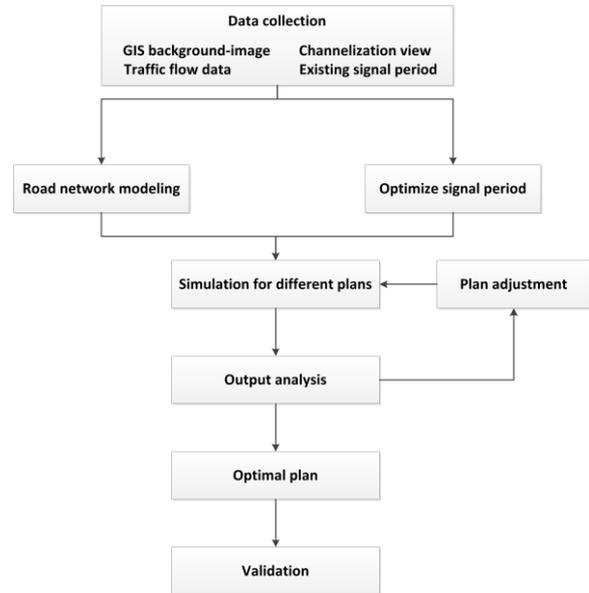

Fig. 3 Main functions and operational process of FLOWSIM.

## 3 Fuzzy logic based traffic modeling theory

Fuzzy logic introduces the uncertainty into simulation process to describe the driving behavior. A set of new variables can be used to express the desire of driver by changing the real observables like speed and headway, into a series of new terms in fuzzy logic like fast and close. Then membership function can be developed to divide one value into several degrees corresponding to reality.

Usually microscopic traffic simulation software contains various models, among which the car-following model and lane-changing model are the most significant. Car-following model decides how the subsequent vehicle will react to the variance of key parameters, like headway distance and velocity difference. Lane-changing model instructs the vehicle to change to the lateral path to get rid of driving pressure or to change direction at intersections. In this section, the theoretical fundamental of the above two models in FLOWSIM is illustrated.

### 3.1 Car-following model

Typically a car-following model has two important variables influencing driving behavior, which are defined as DV (relative speed) and DSSD (the ratio of distance between vehicles to the driver's desired gap). In FLOWSIM these two variables can be divided into fuzzy choices as shown in Table 1.

This fuzzy algorithm describes the following vehicle's response to the changes of relative speed and headway between the leading vehicle. To be mentioned, when following vehicle is at V2 and S4 stage, the following driver still takes no action [3].

Table 1 Fuzzy sets in car-following model in FLOWSIM

| DV (relative speed) | DSSD (distance divergence) | Response (acceleration rate) |
|---|---|---|
| Opening fast (V1) | Much to far (S1) | Strong acceleration |
| Opening (V2) | Too far (S2) | Light acceleration |
| About zero (V3) | Satisfied (S3) | No action |
| Closing (V4) | Too close (S4) | Light deceleration |
| Closing fast (V5) | Much too close (S5) | Strong deceleration |

### 3.2 Lane-changing model

Lane-changing behavior includes moving form offside to the nearside and vice versa. However, these two situations are totally different. If a driver moves from offside to the nearside, he wants to slow down and reduce the pressure form fast moving vehicles behind, while from nearside to the offside is to obtain higher speed and get rid of the pressure form slow moving vehicles ahead.

In FLOWSIM there are two variables for lane-changing behavior. The LCN (lane change to the nearside) is decided by 'pressure form gear' and 'gap satisfaction' in the nearside lane, while 'overtaking benefit' and 'opportunity' for LCO (lane change to offside). Though lane-changing model should be more complicated for its stochastic feature, only three sets of indexes are ruled for simplicity. Table 2 and Table 3 show the fuzzy model for LCN and LCO in FLOWSIM.

Table 2 Fuzzy sets for LCN model

| Pressure form gear | Gap satisfaction | Intention of LCN |
|---|---|---|
| High (PR1) | High (GS1) | High |
| Medium (PR2) | Medium (GS2) | Medium |
| Low (PR3) | Low (GS3) | Low |

Table 3 Fuzzy sets for LCO model

| Overtaking benefit | opportunity | Intention of LCO |
|---|---|---|
| High (OB1) | Good (OP1) | High |
| Medium (OB2) | Moderate (OP2) | Medium |
| Low (OB3) | Bad (OP3) | Low |

The fuzzy rule of lane-changing model describes a driver's intention for a lane change behavior based on desired speed, destination, and the adjacent traffic conditions. A predetermined rule for LCN is that when the pressure form gear is high and gap satisfaction is low, the driver will have a medium intention to change lane [3].

Car-following and lane-changing behavioral data have been collected to calibrate and validate the fuzzy models. Furthermore, in order to verify the model, a series of tests are implemented. Results show that the standard error between car-following model and other models like GIPPS is in an acceptable range, and the lane-changing model can correspond with field data perfectly [8].

## 4 Software Applications

The applications of microscopic traffic simulation software are ranging from control solutions evaluation, signal plan optimization and driving behavior analysis. For the application in China, there are two aspects to be considered. Firstly, in mid-sized and small-sized cities, mixed traffic flow phenomenon is common, which needs simulation. The second is that the large scale signalized intersections in big cities need advanced management methods. In this section, FLOWSIM applications concerning the above two issues are given with examples.

### 4.1 Mixed traffic flow management

In China, the mixed traffic flow phenomenon is increasingly severe in mid-sized cities, due to the lack of isolation facilities between vehicles and non-motor vehicles. In fact, mixed traffic flow creature can decrease the traffic efficiency, thus should be avoided. However, given the traffic development situation, it would take years to build isolation facilities for all intersections. As a result, the current signal control plans provided by typical traffic simulation software may not fit the field condition, which can cause more delay and lower speed [9].

FLOWSIM has been collecting mixed traffic data in China for years, and the conflicts between vehicles and non-motor vehicles can be described in simulation objectively. So it is considerable to evaluate the control plans in intersections for further reconstruction purpose by FLOWSIM. In this subsection, the simulation of a mixed traffic intersection in Huangshi is carried out as an illustration.

Huangshi is a mid-sized city in Hubei province, where the interference between vehicles and non-motor vehicles is very common in the downtown area, especially at intersections. The intersection of Hubinlu-Tianjinlu is located in the urban center, which has a large number of commute traffic during peak hours. To improve traffic efficiency at this intersection, several control plans have been proposed, between which the adjustments of lane channelization and signal plan are the main differences.

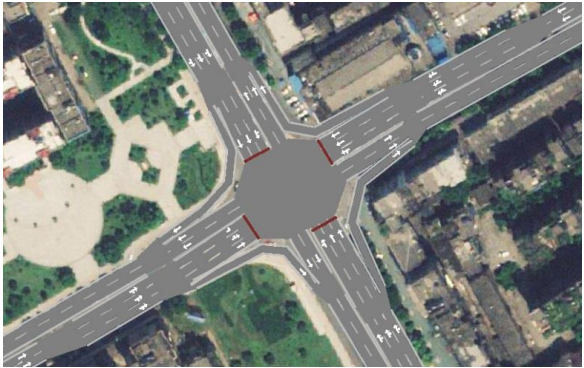

Fig.4 Simulation model of the study intersection.

**Table 4 Phase plans for intersection**

| Plan 1 | Plan 2 | Plan 3 |
| --- | --- | --- |

Plan 1: This plan consists of four signal phases. During each one, the left-turn and through vehicles from the same entrance can approach the intersection, while the right-turn traffic is not controlled by traffic light. Left-turn for non-motor vehicles and pedestrian are prohibited. Instead, they can use the twice crossing method to finish left turning. It is noticed that this plan can avoid most of the interference between vehicles and non-motor vehicles, but the waiting time for green light is longer by contrast.

Plan 2: In this plan, each phase is prepared for the two counter-direction traffic. The left-turn traffic for both vehicles and non-motor vehicles share a protected phase. Also the right-turn traffic is not controlled by signal. Besides, left-turn lane should be adopted. This plan is well known for its commonly application in many cities. However, in mixed traffic flow situation, the conflicts may be severe.

Plan 3: There are only two phases in this plan, which is the simplification of plan 1. In this plan, the green time for each direction is longer than other plans, however, many kinds of conflicts can be observed, which will cause the reduction of travel efficiencies.

The details of three plans are described in Table 4, where the green arrows represent the vehicle traffic flow and the yellow arrows mean the non-motor vehicle and pedestrian flow.

For simulation, the roadway configuration, signal plans and traffic flow data are input into FLOWSIM for model development. The evening peak hour from 5 p.m. to 6 p.m. is selected as simulation period, because the traffic congestion is the severest at that time. After one hour simulation for three plans, key traffic flow parameters such as road density, queue length, average speed, average delay and traffic flow rate can be recorded for each plan. The comparisons of average speed and average delay are shown in Fig. 5.

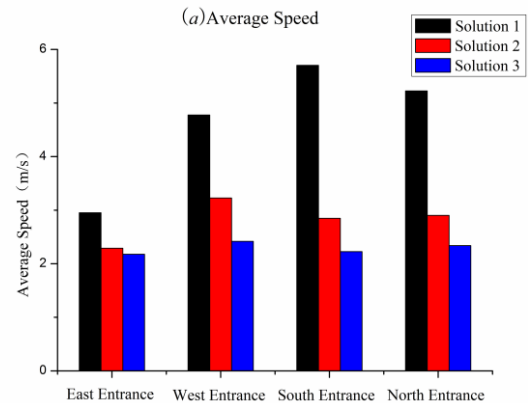

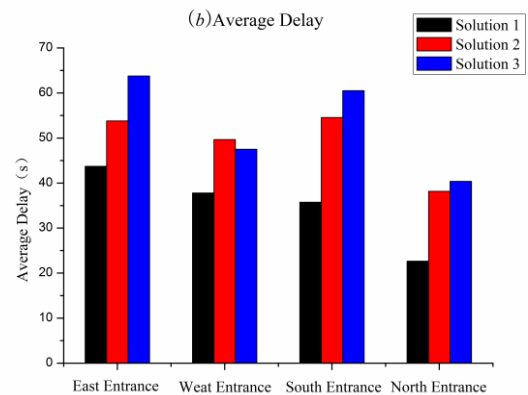

Fig. 5 Comparisons between solutions for (a) speed and (b) delay.

It is clear that plan 1 performs better than others, due to the well-organized mixed traffic flow. With few conflicts, vehicles can reduce their stop times and increase travel speed. Simulation results indicate that even though there are no isolation facilities at intersection, the interference can still be avoided by advanced management methods. It is also concluded that on mixed traffic flow conditions, the control plan should be reconsidered with the mixed traffic interference. As a result, plan 1 can be selected as the reconstruction solution. Also, the twice crossing method for non-motor vehicles to turn left is proved to be effective, because this solution can reduce the conflicts between left-turn vehicles and left-turn non-motor vehicles from the opposite direction [10].

### 4.2 Intersection management assessment

For large cities, real time traffic management is becoming critical because traffic authorities must act quickly in response to traffic disorders. Except for surveillance cameras, other means should be introduced to monitor the traffic management quality, especially at intersections

Dealing with this issue, however, there are few studies focusing on the dynamic intersection management assessment. Management assessment is the evaluation of traffic system performance based on real time traffic data. As a method, the real time simulation results of an intersection can act as criteria for traffic management authorities to assess the working performance and discover the traffic disorders.

In urban areas most intersections are isolated signalized with fixed period during the peak hours. Hence this paper gives an example of ZhongGuanCun 1st Bridge in Beijing as an application demonstration.

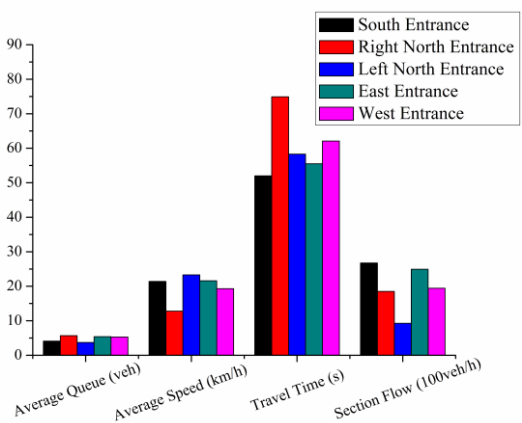

Fig. 6 Statistics at ZhongGuanCun 1st Bridge by FLOWSIM.

The study intersection is in an important commercial and recreational area, with serious congestions every day. With a 15-minute interval traffic data input, the real time simulation can be carried out four times per hour. By simulation at the fastest speed, FLOWSIM can give the results of a 15-minute simulation in less than 2 minutes (at a speed of 7.6 simulation seconds per real second). Because not all the variables can be measured on spot, we take the average queue length, average speed, travel time, and roadway section flow to assess the intersection management quality, shown in Fig. 6.

By simulation every 15 minute, we can compare the dynamic traffic condition of ZhongGuanCun 1st Bridge with simulation results. To be mentioned, the real traffic data can be obtained by loop detectors displaced under road pavement, from which the average speed, traffic volume and occupancy can be archived. If the difference between observed data and simulation result is greater than a predetermined boundary, special actions should be taken to find out the reason. Then the intersection management plan should be changed to special solutions.

In summary, by real time simulation method, traffic management authorities can evaluate the performance of real time traffic system and then implement related solutions.

## 5 Summary and further study

This paper introduces the microscopic simulation software FLOWSIM by giving its fundamental theory and two applications. Work so far has shown that FLOWSIM can be successfully used to describe driver's behavior in China. The theoretical basis is fuzzy logic by considering the drivers decisions as fuzzy ones. Applications like mixed traffic flow management and signalized intersection management assessment are demonstrated. For mid-sized cities, the intersection control plans can be evaluated by FLOWSIM, considering the interference between vehicles and non-motor vehicles. And for large cities, a new management method for traffic authorities is raised, which can assess the real time traffic condition by dynamic simulation.

In this paper, the intersections are simply considered as isolated ones. However, the results may change when take the adjacent intersections into account. Furthermore, it is urgent to introduce the green wave strategy into urban traffic management. So in further study, regional simulation should be investigated both in mid-sized cities and large cities. Besides, regional feedback control theory is another research objective, for the purpose of protecting the urban center.